\documentstyle{article}

\def\QQc{\renewcommand{\baselinestretch}{1.5}}

\begin {document}
\QQc

\def\fo{\hbox{{1}\kern-.25em\hbox{l}}}
\def\fnote#1#2{\begingroup\def\thefootnote{#1}\footnote{#2}\addtocounter
{footnote}{-1}\endgroup}

\def\Cs{$^{133}$Cs}
\def\Tl{$^{205}$Tl}
\def\Hg{$^{199}$Hg}
\def\Xe{$^{129}$Xe}
\def\TlF{$^{205}$TlF}
\def\ecm{e~{\rm cm}}
\def\GF2{\frac{G_F}{\sqrt{2}}}
\def\2GF{\frac{\sqrt{2}}{G_F}}
\def\mse{m_{\tilde{e}}}
\def\mp{m_{\tilde{\gamma}}}
\def\mg{m_{\tilde{\lambda}}}
\def\mq{m_{\tilde{q}}}
\def\phip{\phi_{\tilde{\gamma}}}
\def\phig{\phi_{\tilde{\lambda}}}
\def\sinep{{\rm sin}(\phi_{A_e} - \phi_{\tilde{\gamma}})}
\def\sinqg{{\rm sin}(\phi_{A_q} - \phi_{\tilde{\lambda}})}

\def\100GeV{\left(\frac{100~{\rm GeV}}{M}\right)^2}

\def\dlrmu{\stackrel{\leftrightarrow}{\partial^{\mu}}}
\def\dlrmd{\stackrel{\leftrightarrow}{\partial_{\mu}}}

\begin{titlepage}
\begin{flushright}
\large
UTTG-06-92

\end{flushright}
\huge
\vspace{0.4in}
\renewcommand{\thefootnote}{\fnsymbol{footnote}}
\begin{center}
Bounds on Microscopic Physics from P and T Violation
in Atoms and Molecules\footnote{\normalsize Research
supported in part
by the Robert A. Welch Foundation and NSF Grant PHY 9009850.}\\
\vspace{0.4in}
\QQc
\large
Willy Fischler, Sonia Paban\fnote{\dagger}{\normalsize
paban@utaphy.bitnet}, and
Scott Thomas\fnote{\ddagger}{\normalsize thomas@utaphy.bitnet}\\
Theory Group\\
Department of Physics\\
University of Texas\\
Austin, Texas 78712\\

\vspace{0.2in}

\vspace{0.25in}

Abstract

\end{center}

\large

Atomic and molecular electric dipole moments are calculated within
the minimal supersymmetric standard model.
Present experiments already provide strong bounds on
the combination of phases
responsible for the dipole moments of the neutron and closed shell atoms.
For a supersymmetry breaking scale of 100 GeV, these phases must be
smaller than $ \sim 10^{-2}$.

\large

\end{titlepage}

\large

The experimental bounds on electric dipole moments (edm's)
of the neutron, atoms, and molecules, are reaching a level of
precision that makes them useful probes of P and T violation
that may originate at scales beyond the standard model.
Aside from the QCD vacuum angle, the only source of
T (or CP) violation in the standard model is the
phase in the quark mass matrix.
This generates edm's orders of magnitude beyond experimental
reach \cite{smedm}.
Extensions of the standard model however generally have
CP violating phases that produce edm's of the
order of, or larger than, present
experimental bounds.
A finite $\bar{\theta}_{QCD}$ that saturates the bound from the
neutron edm \cite{neutron}, would also produce atomic edm's
of the same order.
In this paper the main sources for atomic
and molecular
edm's in the minimal supersymmetric extension of the standard model
will be identified.

In order to identify the microscopic sources
of P and T violation
responsible for the edm's, it is instructive to present the
analysis starting from the largest relevant scale, namely
the atomic scale.
The next step is then to proceed down to the nuclear scale.
This will enable finally a discussion of the origin
of P and T violation at the subnuclear level.

At the atomic scale an atom or molecule is a composite system of
electrons and nuclei. These constituents contribute to the
edm via T and P odd electromagnetic moments and local
non-electromagnetic interactions.
Some of the electromagnetic moments are suppressed.
In the nonrelativistic limit, the contributions from the electron and
nuclear edm's vanish.
This is known in the literature as Schiff's theorem \cite{schiff}.
The electron edm can contribute though in heavy atoms where the
electrons are relativistic \cite{erel,electron}.
But in closed shell atoms and molecules
with paired electron spins, the electron edm
contributes only through hyperfine interactions with
the nucleus, and is therefore suppressed \cite{ehf}.
Schiff's theorem does not apply to higher
moments such as the nuclear magnetic quadrupole moment.
In addition there is another T and P odd moment known as Schiff's
moment, which is proportional to
the offset of electric charge
and dipole distributions of the nucleus.
It leads to a local electromagnetic coupling
of the electron to the nucleus \cite{schiff}.

At the nuclear scale
all of the contributions discussed above are contained in the following
Hamiltonian
$$
H = \frac{1}{2}d_e \bar{e} \sigma_{\mu \nu} i \gamma_5 e F^{\mu \nu}
+   \frac{1}{2}d_N \bar{N} \sigma_{\mu \nu} i \gamma_5 N F^{\mu \nu}
+ C_{PS,S}^{en} \GF2
     (\bar{e} i \gamma_5 e) ( \bar{N} N)
$$
$$
+ C_{S,PS}^{en} \GF2
     (\bar{e} e) (\bar{N} i \gamma_5 N)
+ C_{T,PT}^{en} \GF2
     (\bar{e} \sigma_{\mu \nu} e) (\bar{N} \sigma^{\mu \nu} i \gamma_5 N)
$$
$$
+ Q_{V,PV}^{en}
     (\bar{e} \gamma_{\mu} e) ( \bar{N} \dlrmu \gamma_5 N )
+ Q_{PV,V}^{en}
     (\bar{e} \dlrmd \gamma_5 e ) ( \bar{N} \gamma^{\mu} N )
$$
\begin{equation}
+ C_{S,PS}^{nn} \GF2
     ( \bar{N} N) ( \bar{N} i \gamma_5 N )
+ Q_{V,PV}^{nn}
     (\bar{N} \gamma_{\mu} N) ( \bar{N} \dlrmu \gamma_5 N)
\end{equation}
where $N=(p,n)$ and isospin violation is ignored.
The first and second terms are the electron and nucleon edm's
respectively.
The remaining terms are local interactions among the electrons
and nucleons.
The nucleon edm and nucleon-nucleon couplings contribute
to the Schiff and magnetic quadrupole moments \cite{nn}.
The electron-nucleon
couplings of course contribute to the electron-nucleus
coupling \cite{en}.
Calculations of the contributions of the terms in $H$ to the
edm's considered below may be found in
Ref.s [5,7-11]
and
are summarized in Table 1.
The contributions from
$Q_{V,PV}^{en}$,
$Q_{PV,V}^{en}$, and $Q_{V,PV}^{nn}$ have not been considered
in these references although $Q_{V,PV}^{en}$ does contribute directly
to the Schiff moment.
These derivative operators have the same nonrelativistic limit as
some of the nonderivative operators.
This allows a rough estimate of the corresponding contributions
to the edm's.

In this paper the microscopic origin of the operators in
(1) will be calculated within the context of the
minimal supersymmetric extension
of the standard model \cite{a}.
In addition to the KM like phases, this model has
other CP violating phases
in the superpotential and soft supersymmetry breaking
terms.
The superpotential contributes a phase from the term
$ | \mu | e^{i \phi_{\mu}} {\hat{H}}_1 {\hat{H}}_2$.
The phases in the soft breaking terms arise from the gaugino
masses,
$-\frac{1}{2} |m_{\tilde a}| e^{i \phi_{\tilde a}}
  \lambda_{\tilde a} \lambda_{\tilde a} ~ + ~ {\rm h.c.}$,
and the mass parameters coupling left and right handed squarks
(sleptons),
$-f_L^* m_f |A_f| e^{i \phi_{A_f}} f_R ~ + ~ {\rm h.c.}$,
where $m_f$ is the corresponding quark (lepton) mass and
$A_f$ is defined by
\begin{equation}
m_f A_f = m_f \left( \frac{A_F}{h_f} + \mu^{ \ast}
     \frac{v_1}{v_2} \right)
\end{equation}
$A_F$ is the scalar trilinear soft breaking coupling and $h_f$
the Yukawa coupling.
The importance of these new phases is that P and T odd
operators can be
generated irrespective of generational mixing.\fnote{1}{\normalsize
If the hidden sector is Polonyi and the
electroweak gaugino masses vanish at tree level, the supersymmetric
phases can be rotated away.}

The supersymmetric origin of the terms
in the Hamiltonian (1) will now be considered.

1. The electron edm is a chirally violating operator of
effective dimension 6.
It arises at the one loop level from
electroweak gaugino exchange as shown in Fig. 1.
The sum of all such diagrams contains several unknown masses and
phases \cite{Nath}.
In order to display the order of magnitude,
only the photino contribution will be considered \cite{eEDM},
\begin{equation}
d_e = - e \frac{\alpha}{ 24 \pi} \frac{m_e |A_e|}{\mp^3}
      {\rm sin}(\phi_{A_e} - \phip) f(x) ~ +~ \cdots
\end{equation}
where
$x=\mse^2 / \mp^2$, $f(x)$ is a loop function of order
unity, $f(1)=1$ \cite{eEDM}, and $ + \cdots$ represents the
contributions from the other electroweak gauginos.
Numerically, for $ \mse = \mp $ Eq. (3) gives \cite{eEDM}
\begin{equation}
d_e \simeq -1 \times 10^{-25}
    \frac{( 100~ {\rm GeV})^2}{\mp^3/ |A_e|}
    {\rm sin}(\phi_{A_e} - \phip)
    ~ \ecm ~ + ~ \cdots
\end{equation}

2. The lowest dimension operators which potentially
contribute to the nucleon
edm are the light quark electric and
chromoelectric dipole moments \cite{susy,Arnowitt},
and Weinberg's 3-gluon operator \cite{Weinberg,3g}
$$
\begin{array}{l}
{\cal O}_1 = \frac{1}{2} \bar{q} \sigma_{\mu \nu} i \gamma_5 q
   F^{\mu \nu}  \\
{\cal O}_2 = \frac{1}{2} \bar{q} \sigma_{\mu \nu} i \gamma_5
   {\rm T}_a q G^{\mu \nu}_a \\
{\cal O}_3 = \frac{1}{3} f_{abc} \tilde{G}_{a \mu \nu}
     G_b^{\mu \rho} G_{c \rho}^{\nu}
\end{array}
$$
where $\tilde{G}_{a \mu \nu} = \frac{1}{2}
\epsilon_{\mu \nu \rho \sigma} G_a^{\rho \sigma}$
and ${\rm T}_a$ are the SU(3)$_C$ generators.
As for the electron edm
the quark dipole moments are effectively dimension-6.
The operators ${\cal O}_1$ and ${\cal O}_2$
are generated at the one loop level
through gluino and electroweak gaugino exchange \cite{Arnowitt}.
The sum of all such diagrams is again a function of several masses
and phases.
To establish the order of magnitude only the gluino
and photino exchange will be considered
\begin{equation}
C_1(\mu)= \frac{e Q_q}{24 \pi} m_q |A_q| \left(
   \frac{\alpha_s}{\mg^3} {\rm sin}(\phi_{A_q} - \phig) \frac{4}{3} f(y)
 + \frac{\alpha Q^2_q}{\mp^3} {\rm sin}(\phi_{A_q} - \phip)  f(z)
 +  \cdots    \right)  \zeta_1
\end{equation}
\begin{equation}
C_2(\mu)= - \frac{g_s}{24 \pi} m_q |A_q| \left(
  \frac{\alpha_s}{\mg^3} {\rm sin}(\phi_{A_q} - \phig)
    ( h(y) + \frac{1}{6} f(y) )
 - \frac{\alpha Q_q^2}{\mp^3} {\rm sin}( \phi_{A_q} - \phip) f(z)
 +  \cdots  \right) \zeta_2
\end{equation}
where $Q_q$ is the quark charge,
$y=\mq^2 / \mg^2$,$~z=\mq^2 / \mp^2 $,
$f$ is the same function that appears in Eq. (3),
and $h(1)=1$ \cite{Arnowitt}.
$\zeta_i$ are renormalization group corrections for the evolution
of the operators from the supersymmetric scale, $M$, to the
hadronic scale, $\mu$.
Using the known anomalous dimensions \cite{run} with
$\alpha_s(M) = .1$ and $\alpha_s(\mu) = 4 \pi / 6$ \cite{Weinberg}
gives $\zeta_1 \simeq 1.6$, and $ \zeta_2 \simeq 3.6$.
Note that the photino contribution is down by
$ \sim \alpha \alpha_s^{-1}$
compared with the gluino.
Barring any cancellation among the phases, and {\it if all the
mass parameters are of the
same order}, the gluino contribution will dominate the quark dipole
moments.

The 3-gluon operator is generated at the two loop level.
The largest
contributions come from integrating out the
chromoelectric dipole moments of quarks with mass
$m_Q > \Lambda_{QCD}$ \cite{3g,run},
\begin{equation}
C_3(\mu)= \frac{\alpha_s(m_Q)}{8 \pi} \frac{C_2(m_Q)}{m_Q} \zeta_3
\end{equation}
Keeping only the charm quark as the dominant contribution gives
$\zeta_3 \simeq .38$.

An estimate of the edm requires an evaluation of the
matrix elements of these operators on the nucleon.
There is however no systematic approximation scheme to reliably
calculate hadronic matrix elements of this type.
Here, some empirical rules known as ``naive dimensional
analysis'' that keep track of factors of $4 \pi$ and mass scales
 \cite{Weinberg,NDA} will be used.
This is the most uncertain aspect of the entire analysis.
Even so, there is no physical reason to expect a substantial
suppression or enhancement as compared to estimates given by these
rules.
Using these rules, the nucleon edm is
\begin{equation}
d_N \simeq C_1(\mu) + \frac{e}{4 \pi} C_2(\mu) +
      \frac{e}{4 \pi} \Lambda_{\chi} C_3(\mu)
\end{equation}
where $\Lambda_{\chi} \simeq 4 \pi f_{\pi}$ is the chiral
symmetry breaking scale.
Numerically, keeping only the gluino contributions with
$m_{\tilde q} = m_{\tilde{\lambda}} =
m_{\tilde{\gamma}} \equiv \tilde{m}$,
and $A_u = A_d = A_c \equiv A_q$, Eq. (8) then gives
\begin{equation}
|d_N| \simeq \left(
  2.2 + 1.1 + .1 \right) ~
  \frac{(100 ~ {\rm GeV})^2}{\tilde{m}^3 / |A_q|}
  \sinqg ~
  10^{-23} ~ \ecm
\end{equation}
where the the terms on the right side come from
${\cal O}_1$, ${\cal O}_2$, and ${\cal O}_3$ respectively.

3. The nucleon-nucleon couplings are just the T odd component
of the nuclear force.
Following Weinberg's analysis of nuclear forces \cite{nuc}, it is
useful to think of these as arising from the T odd exchange of
low lying mesons and direct contact interactions.
For illustrative purposes only
the exchange of the lightest meson, the pion,
will be considered; and
an estimate of the T odd pion-nucleon
coupling,
$\bar{g}_{\pi N N} \bar{N} \pi N$,
arising from the microscopic physics will be made.
The leading contribution will come from the light
quark chromoelectric dipole moment.
Using ``naive dimensional analysis'' the coupling is
\begin{equation}
\bar{g}_{\pi N N} \simeq g_{\pi N N} \frac{\Lambda_{\chi}}{4 \pi}
 C_2(\mu)
\label{pnn}
\end{equation}
where $g_{\pi N N}$ is the usual pseudoscalar pion-nucleon coupling
constant.
Integrating out the pion, the
contribution to the nonderivative nucleon-nucleon coupling is
\begin{equation}
C^{nn}_{S,PS} \sim {\bar{g}}_{\pi N N} g_{\pi N N }
\frac{1}{{m_{\pi}}^2} \frac{\sqrt{2}}{G_F}
\end{equation}
Numerically, keeping only the gluino contribution with
$\mg = \mq$, Eqs. (6), (10), and (11) give
\begin{equation}
C_{S,PS}^{nn} \simeq .7 ~
  \frac{(100~ {\rm GeV})^2}{\mq^3/ |A_q|} \sinqg
\end{equation}
The exchange of heavier mesons will also contribute to
this coupling and the derivative
nucleon-nucleon coupling.
This is not expected to substantially change the estimates
given below based solely on the nonderivative coupling from pion
exchange.

4. The electron-nucleon couplings in (1) arise from two classes
of operators.
The first operators involve an electromagnetic infrared enhancement and
are suppressed by two powers of a heavy mass.
The second class of operators are local at the atomic
scale and suppressed by four powers of a heavy mass.

The first class of operators arise from the effective nucleon-photon
couplings $\bar{N} N F_{\mu \nu} \tilde{F}^{\mu \nu}$ and
$\bar{N} i \gamma_5 N F_{\mu \nu} F^{\mu \nu}$.
The two photons couple to electrons to produce an
effective electron-nucleon interaction.
Integrating out the photons gives an infrared enhancement,
cutoff by the electron mass.
One way in which such an effective interaction arises is through
the pion pole as shown in Fig. 2, where the CP violation
occurs in the pion-nucleon coupling.
This contributes to the electron-nucleon pseudoscalar-scalar
coupling
\begin{equation}
C^{en}_{PS,S} \sim g_{\pi ee} {\bar{g}}_{\pi NN} \frac{1}{m_{\pi}^2}
\frac{\sqrt{2}}{G_F}
\label{pie}
\end{equation}
where $\bar{g}_{\pi N N}$ is given in (10), $g_{\pi ee}$
is the effective pion-electron coupling computed in
Ref. \cite{choi}
\begin{equation}
g_{\pi ee} \sim - \frac{3 {\alpha}^2}{2 {\pi}^2} \frac{m_e}{f_{\pi}}
\ln \left( \frac{\Lambda_{\chi}}{m_e} \right)
\end{equation}
and the approximation that the electron is on
shell has been used.
Note that the effective electron-nucleon interaction
is suppressed by only two powers of a heavy mass
(from the light quark chromoelectric dipole moment).
A similar diagram with the CP violation in the vertex
$\pi^0 F_{\mu \nu} F^{\mu \nu}$,
contributes to the scalar-pseudoscalar coupling.
The edm's in Table 1. are however somewhat less sensitive to
this coupling.

The nucleon-photon couplings can also arise from the
following microscopic operators which are not suppressed by
a light quark mass
$$
\begin{array}{l}
{\cal O}_4 = G_{a \mu \nu} G_{a}^{\mu \nu} F_{\rho \sigma}
    \tilde{F}^{\rho \sigma}  \\
{\cal O}_5 = G_{a \mu \nu} \tilde{G}_{a}^{\mu \nu}
    F_{\rho \sigma} F^{\rho \sigma}
\end{array}
$$
These operators arise at the two loop level.
The dominant contribution to ${\cal O}_4$ comes from integrating
out the electric dipole moment of a quark with mass
$m_Q > \Lambda_{QCD}$, as
shown in Fig. 3.
This involves an infrared enhancement, cutoff of by the quark
mass.
The coefficient is therefore suppressed by only two powers of
a heavy mass.
Similar operators have been considered in Ref. \cite{ab}.
Here,
\begin{equation}
C_{4} \simeq \frac{ e {g_s}^2 Q_Q}{256 {\pi}^2 {m_Q}^3} C_1(m_Q)
\end{equation}
Using ``naive dimensional analysis'' then
\begin{equation}
C^{en}_{PS,S} \simeq -6 \frac{\alpha}{\pi} m_e \Lambda_{\chi}
\ln \left( \frac{m_Q}{m_e} \right)
\frac{\sqrt{2}}{G_F} C_{4}
\end{equation}
A similar discussion applies to ${\cal O}_5$ with the
chromoelectric dipole moment of heavy quarks.
All these operators scale like $(m_Q M)^{-2}$.

It should be noted that the infrared enhancement from the
photon-electron loop comes from momenta $\sim m_e$.
Treating the resulting interactions as local is somewhat
dubious for heavy atoms for which $p_e \sim Z \alpha m_e$.
This is not expected to significantly alter the estimates
of the edm's given below.
Among the first class of operators, the estimate
in Eq. (13) turns out to be numerically most important,
\begin{equation}
C^{en}_{PS,S} \sim 2 \times 10^{-8}
   \frac{(100~{\rm GeV})^2}{\mq^3 / |A_q|}
   {\rm sin}(\phi_{A_q} - \phig)
\end{equation}

The second class of local operators that lead to
electron-nucleon couplings include
$$
\begin{array}{ll}
{\cal O}_6 = \bar{e} i \gamma_5 e \bar{q} q     & ~~
{\cal O}_7 = \bar{e} e \bar{q} i \gamma_5 q     \\
{\cal O}_8 = \bar{e} \sigma_{\mu \nu} e \bar{q} \sigma^{\mu \nu}
    i \gamma_5 q   & ~~
{\cal O}_9 = \bar{e} i \gamma_5 e G_{a \mu \nu}
G_a^{\mu \nu}      \\
{\cal O}_{10} = \bar{e} e G_{a \mu \nu} \tilde{G}_a^{\mu \nu} & ~~
{\cal O}_{11} = \bar{e} \gamma_{\mu} e
     \bar{q} \dlrmu \gamma_5 q       \\
{\cal O}_{12} = \bar{e} \dlrmd \gamma_5 e
     \bar{q} \gamma^{\mu} q \\
\end{array}
$$
Because of the chiral properties,
all these operators are effectively dimension 8.
The operators ${\cal O}_6, {\cal O}_7$, and ${\cal O}_8$ are
generated, after Fierz reordering, by box
diagrams\fnote{2}{\normalsize Tree level Higgs exchange is CP
conserving in the minimal supersymmetric standard model and will
not contribute to the electron-nucleon couplings.}
of the type shown in Fig. 4.
Neglecting running corrections, the coefficients are then related by
$C_6 = C_7 = \frac{1}{2} C_8 \equiv C$.
A typical contribution is of order
\begin{equation}
C \sim \frac{\alpha}{4 \pi} G_F
   \frac{m_e m_q |\mu|}{{\rm sin} 2 \beta~ \mp^3}
  {\rm sin}(\phi_{\mu} - \phip)
\end{equation}
where $\tan{\beta}= {v_2}/{v_1}$.
With all the supersymmetric mass parameters of order $M$, this
scale like $G_F M^{-2}$.

The chiral suppression of a light quark mass is avoided in the
operators ${\cal O}_9$ and ${\cal O}_{10}$.
These can arise from the box diagrams of Fig. 4 with the light quarks
replaced by heavy quarks, $Q$.
The heavy quarks are then integrated out as shown in Fig. 5.
In the limit $M >> m_Q$,
$$
C_9 = \sum_{Q} -\frac{2}{3} \frac{\alpha_s}{4 \pi} \frac{C_6(m_Q)}{m_Q}
$$
\begin{equation}
C_{10} = \sum_{Q} -\frac{\alpha_s}{4 \pi} \frac{C_7(m_Q)}{m_Q}
\end{equation}
There are no dimension-8 operators without a light
quark mass suppression that give a
tensor-pseudotensor Lorentz structure.
There is however an effective dimension-10 operator,
${\cal O}_{13} = \frac{1}{3} \bar{e} \sigma_{\mu \nu} e i d_{abc}
\tilde{G}_{a \rho \sigma} G_{b}^{\rho \sigma} G_{c}^{\mu \nu}$,
which is generated analogously to ${\cal O}_9$ and
${\cal O}_{10}$, but with at least three gluons on
the heavy quark loop.
With the rules of ``naive dimensional analysis'' this operator is
not much suppressed compared with ${\cal O}_9$ and
${\cal O}_{10}$.
It is therefore included in the estimate below.
Using ``naive dimensional analysis'' the contributions of
${\cal O}_6$ through ${\cal O}_{10}$, and ${\cal O}_{13}$,
to the nonderivative
electron-nucleon couplings are
$$
C_{PS,S}^{en} = \2GF \left( C_6(\mu)
    - \sum_{Q} \frac{2}{3} \frac{\alpha_s}{4 \pi}
      \frac{\Lambda_{\chi}}{m_Q}  C_6(m_Q) \right)
$$
$$
C_{S,PS}^{en} = \2GF \left( C_7(\mu)
    - \sum_{Q} \frac{\alpha_s}{4 \pi}
      \frac{\Lambda_{\chi}}{m_Q}  C_7(m_Q) \right)
$$
\begin{equation}
C_{T,PT}^{en} = \2GF \left(  C_8(\mu)
    - \sum_{Q} \frac{1}{24} \left( \frac{g_s}{4 \pi} \right)^3
    \left( \frac{\Lambda_{\chi}}{m_Q} \right)^3
    C_8(m_Q) \right)
\end{equation}
where again running corrections have been neglected.

The operators ${\cal O}_{11}$ and $ {\cal O}_{12}$ are
generated by box diagrams of the type
shown in Fig. 6a,
$$
C_{11} \sim \frac{\alpha}{ 4 \pi} e^2 Q_q^2
    \frac{m_q |A_q|}{ \mp^5} {\rm sin}(\phip - \phi_{A_q})
$$
\begin{equation}
C_{12} \sim \frac{\alpha}{ 4 \pi} e^2 Q_q^2
    \frac{m_e |A_e|}{ \mp^5} {\rm sin}(\phip - \phi_{A_e})
\end{equation}
There are also contributions that result from a weak interaction
between the electron(quark) and the weak-edm of the quark(electron)
as shown in Fig. 6b.
$$
C_{11} = \pm (1-4 |Q_q| {\rm sin}^2 \theta_w)(1-4 {\rm sin}^2 \theta_w)
   \GF2 \frac{C_1(\mu)}{e}
$$
\begin{equation}
C_{12} = \pm (1-4 |Q_q| {\rm sin}^2 \theta_w) {\rm cot} \theta_w
   \GF2 \frac{d_e^{(W)}}{e}
\end{equation}
where $\pm$ refers to up or down type quarks,
$d_e^{(W)}$ is the electron edm arising from wino exchange,
and $C_1(\mu)$ is given above.
These operators scale like either $G_F M^{-2}$ or $M^{-4}$.
Using ``naive dimensional analysis''
the contribution to the derivative electron-nucleon couplings in
(1) are
$$
Q_{V,PV}^{en} =  C_{11}
$$
\begin{equation}
Q_{PV,V}^{en} = C_{12}
\end{equation}
All of the local operators turn out to be numerically
somewhat less important than the estimate in Eq. (13).

The atomic and molecular edm's can now be estimated as functions
of the microscopic parameters.
In order to identify the most important effects at the atomic scale,
the contributions from the electron edm, nucleon edm, nonderivative
nucleon-nucleon coupling, and the largest electron-nucleon coupling
(i.e. from Eq. (13)) will be retained.
Since an edm is proportional to spin, the leading contributions
in open and closed shell atoms are different.
For open shell atoms we consider \Cs~ and \Tl~
since good experimental bounds are
available \cite{amherst,berkeley}.
Putting together the results cataloged above with the results of
the atomic and nuclear calculations from Table 1.,
$$
d_{\rm Cs} \simeq ( ~
  - 1.2 ~ \sinep
  + 2 \times 10^{-3} \sinqg
  + 1 \times 10^{-1} \sinqg
$$
\begin{equation}
  + 1 \times 10^{-3} \sinqg
    ~ )
  \100GeV ~
  10^{-23} ~\ecm
\end{equation}
$$
d_{\rm Tl} \simeq (~
    6 ~ \sinep
  + 2 \times 10^{-4} \sinqg
  + 2 \times 10^{-3} \sinqg
$$
\begin{equation}
  - 1 \times 10^{-2} \sinqg
    ~ )
  \100GeV ~
  10^{-23} ~\ecm
\end{equation}
where, for simplicity of notation, all the supersymmetric mass
parameters have been assumed to be of order $M$.
The quantities on the right hand side arise respectively from
the electron edm, nucleon edm, nonderivative nucleon-nucleon coupling,
and pseudoscalar-scalar electron-nucleon coupling.
In both cases, the electron edm gives the dominant contribution
since it is enhanced in heavy open shell atoms.
For \Cs, with a nuclear quadrupole moment, the nucleon-nucleon
coupling is only a factor $\sim$ 10 less important than the
electron edm (again assuming all the masses and phases are the
same order).
The present experimental bounds
are \cite{amherst,berkeley}\fnote{3}{\normalsize Unless
stated otherwise, the
experimental bounds given here are the sum of the reported
measurement and experimental error. To date, all measurements are
consistent with zero.}
$$
|d_{\rm Cs}| < 7.2 \times 10^{-24} ~\ecm
$$
$$
|d_{\rm Tl}| < 6.6 \times 10^{-24} ~\ecm
$$
The \Tl~ result gives a bound on the phases contributing to the
electron edm of
$$
\sinep \100GeV < .1
$$

For closed shell atoms, good experimental bounds are available for
\Xe~ and \Hg.
Combining our results with those from Table 1.,
$$
d_{\rm Xe} \simeq ( ~
    8 \times 10^{-3} \sinep
  - 7 \times 10^{-2} \sinqg
  + 3 ~ \sinqg
$$
\begin{equation}
  + 2 \times 10^{-4} \sinqg
     ~ )
  \100GeV ~
  10^{-26} ~\ecm
\end{equation}
$$
d_{\rm Hg} \simeq ( ~
  - 1.2 \times 10^{-2} \sinep
  - 3 \times 10^{-2} \sinqg
  + 4 ~ \sinqg
$$
\begin{equation}
  + 2 \times 10^{-4} \sinqg
     ~ )
  \100GeV ~
  10^{-25} ~\ecm
\end{equation}
The electron edm is here suppressed since the electron spins are paired.
The leading contribution comes from the nucleon-nucleon coupling.
The present experimental bounds
are [26-29],
$$
|d_{\rm Xe}| < 1.4 \times 10^{-26} ~\ecm
$$
$$
|d_{\rm Hg}| < 3 \times 10^{-27} ~\ecm ~~(95 \% ~ {\rm C.L.})
$$
The \Hg~ result gives a bound on the phases
contributing to the light quark chromoelectric dipole
moment,
$$
\sinqg \100GeV < .008
$$
It should be noted that, due to the uncertainties in the hadronic
matrix elements and nuclear calculations \cite{nn}, this
bound is much more uncertain than that from the electron
edm given above.

At present, the best bound on a molecular edm is for \TlF.
Again combining our results with Table 1.,
$$
d_{\rm TlF} \simeq ( ~
  - 8 \times 10^{-2} \sinep
  - 1.5 \times 10^{-1} \sinqg
  + \sinqg
$$
\begin{equation}
  - 1 \times 10^{-3} \sinqg
    ~ )
  \100GeV ~
  10^{-22} ~\ecm
\end{equation}
The leading contribution again comes from the nucleon-nucleon
coupling since the electron spins are paired.
The current experimental bound \cite{yaleA,yalelett}
$$
|d_{\rm TlF}| < 4.6 \times 10^{-23} ~\ecm
$$
does not yet yield a substantial bound on the
phases if $M \simeq 100~{\rm GeV}$.

Because of the different structure of open and closed
shell atoms, the experiments bound
different microscopic CP violating parameters.
Open shell atoms are sensitive to phases in the weak
gaugino sector contributing to the electron edm.
Closed shell atoms and molecules with paired electron spin
get the leading
contribution from nuclear effects.
This provides a stringent bound on the gluino-squark
phases contributing to
the light quark chromoelectric dipole moment.
For comparison, the neutron edm receives contributions from
both the light quark electric and chromoelectric dipole
moments \cite{Arnowitt}.
The experimental bound \cite{neutron} of
$|d_n| < 8 \times 10^{-26} ~\ecm$,
with the estimate (9), gives
$$
\sinqg \100GeV < .003
$$
This is of the same order as the bound from \Hg.
Taken together, the experimental bounds provide information
on different combinations of CP violating phases.
Other sources of CP violation would give rise to different
patterns of edm's \cite{higgs}.
For example, a nonzero $\bar{\theta}_{QCD}$ would contribute to the edm's
mainly through the T odd pion-nucleon
coupling \cite{nn,thetanuc}.
This would give a definite relation among the edm's of
open and closed shell atoms and the neutron.

The prospects for improving the present measurements are encouraging.
The ultimate sensitivity of the current \Hg~ experiment
is expected to reach the level of $3 \times 10^{-28}~\ecm$
\cite{fortson}.
New techniques may allow further improvement in sensitivity for
open shell atoms \cite{dan}.
An unexplored area where the phases
contributing to the light quark chromoelectric dipole moment
could be measured is in
light atoms.
In this case the electrons are nonrelativistic and the primary
contributions are nuclear.
In summary, the experimental study of atomic edm's provides a
promising probe for exploring physics beyond the standard model.

We would like to thank D. Heinzen and V. Kaplunovsky for useful
discussions. We would also like to thank N. Fortson for providing
us with the unpublished bound on \Hg.
This research was supported in part by the Robert A. Welch Foundation
and NSF Grant PHY 9009850.

\newpage

\vspace{1.5in.}

$$
\vspace{.2in}
\begin{array}{c|c|c|c|c|c|}
     \multicolumn{1}{c}{ }
   & \multicolumn{1}{c}{^{133}{\rm Cs}}
   & \multicolumn{1}{c}{^{205}{\rm Tl}}
   & \multicolumn{1}{c}{^{129}{\rm Xe}}
   & \multicolumn{1}{c}{^{199}{\rm Hg}}
   & \multicolumn{1}{c}{^{205}{\rm TlF}}  \\ \cline{2-6}
d/d_e  & 120  &  -600   &  -.8 \times 10^{-3}  & 1.2 \times 10^{-2}
  &  80 \\  \cline{2-6}
d/d_N  & 7 \times 10^{-4}  & 7 \times 10^{-5}
  &  2 \times 10^{-5} & \sim 10^{-4} & .5 \\  \cline{2-6}
d (\ecm)/C_{S,PS}^{nn} & 2 \times 10^{-24} & 3 \times 10^{-26}
  &  5 \times 10^{-26} & 6 \times 10^{-25}
  &  2 \times 10^{-22} \\   \cline{2-6}
d (\ecm)/C_{PS,S}^{en} & 7 \times 10^{-19}  & - 5 \times 10^{-18}
  & 9 \times 10^{-23} & 1 \times 10^{-21}
  & -6 \times 10^{-18} \\  \cline{2-6}
d (\ecm)/C_{T,PT}^{en} & 9 \times 10^{-21} & 5 \times 10^{-21}
  & 5 \times 10^{-21} & -6 \times 10^{-20}
  & 1 \times 10^{-16} \\ \cline{2-6}
d (\ecm)/C_{S,PS}^{en} & & & -1 \times 10^{-23}  & 3 \times 10^{-23}
  & -4 \times 10^{-19} \\ \cline{2-6}
\end{array}
\vspace{.6in}
$$

\begin{description}

\item[Table 1.] Contributions to electric dipole moments from
the Hamiltonian (1).

\end{description}

\vspace{1in.}

\newpage
\begin{center}
{\bf Figure Captions}
\end{center}

\begin{description}

\item[Fig. 1] A typical contribution to $d_e$ from photino exchange.

\item[Fig. 2] The contribution to an effective electron-nucleon coupling
from pion exchange with the T odd pion-nucleon coupling.

\item[Fig. 3] The contribution to the operator
$G_{a \mu \nu} G_{a}^{\mu \nu} F_{\rho \sigma}
   \tilde{F}^{\rho \sigma}$ from the electric dipole moment of
a heavy quark.  Other diagrams related by gauge invariance are
not shown.

\item[Fig. 4] A typical contribution to the electron-quark operators.

\item[Fig. 5] The contribution to the electron-gluon operators
from the electron-heavy quark operators.  Other diagrams related
by gauge invariance are not shown.

\item[Fig. 6] Typical contributions to the operator ${\bar{e}} \dlrmd
\gamma_5 e \bar{q} \gamma^{\mu} q$; (6a) from a box diagram, (6b) from
Z-exchange and the electron electroweak dipole moment.

\end{description}

\end{document}